\documentclass[mathleft]{an}
\usepackage{times}
\usepackage{graphicx}
\usepackage{url}
\usepackage{bm}
\graphicspath{{./fig/}{./png/}}

\newcommand{\Beq}{B_{\rm eq}}
\newcommand{\EQ}{\begin{equation}}
\newcommand{\EN}{\end{equation}}
\newcommand{\EQA}{\begin{eqnarray}}
\newcommand{\ENA}{\end{eqnarray}}

\newcommand{\Eq}[1]{Eq.~(\ref{#1})}
\newcommand{\Eqs}[2]{Eqs.~(\ref{#1}) and~(\ref{#2})}

\newcommand{\Fig}[1]{Fig.~\ref{#1}}

\newcommand{\bra}[1]{\langle #1\rangle}

\newcommand{\meanEMF}{\overline{\vec{\cal E}}}

\newcommand{\meanFF}{\overline{\mbox{\boldmath ${\cal F}$}} {}}

\newcommand{\meanAA}{\overline{\vec{A}}}
\newcommand{\meanBB}{\overline{\vec{B}}}
\newcommand{\meanJJ}{\overline{\vec{J}}}
\newcommand{\meanUU}{\overline{\vec{U}}}

\newcommand{\zzz}{\mbox{\boldmath $z$} {}}

\newcommand{\ggamma}{{\vec{\gamma}}}

\newcommand{\uu}{{\vec{u}}}
\newcommand{\BB}{{\vec{B}}}

\newcommand{\bb}{{\vec{b}}}

\newcommand{\nab}{\mbox{\boldmath $\nabla$} {}}

\newcommand{\sgn}{{\rm sgn}  \, {}}

\newcommand{\dd}{{\rm d} {}}

\def\Rm{R_{\rm m}}

\def\erf{\mbox{\rm erf}}

\def\kf{k_{\rm f}}

\def\etat{\eta_{\rm t}}

\def\alphaK{\alpha_{\rm K}}
\def\alphaM{\alpha_{\rm M}}

\def\half{{\textstyle{1\over2}}}

\newcommand{\yapj}[3]{: #1, {ApJ} {#2}, #3}
\newcommand{\yapjl}[3]{: #1, {ApJ} {#2}, #3}

\newcommand{\yan}[3]{: #1, {AN} {#2}, #3}

\newcommand{\yana}[3]{: #1, {A\&A} {#2}, #3}

\newcommand{\ygafd}[3]{: #1, {GApFD} {#2}, #3}

\newcommand{\yprl}[3]{: #1, {Phys Rev Lett} {#2}, #3}
\newcommand{\ypre}[3]{: #1, {Phys Rev E} {#2}, #3}

\newcommand{\ymn}[3]{: #1, {MNRAS} {#2}, #3}

\newcommand{\ysph}[3]{: #1, {Solar Phys.} {#2}, #3}

\newcommand{\yjour}[4]{: #1, {#2} {#3}, #4}

\newcommand{\ybook}[3]{: #1, {\it #2} (#3)}

\Pagespan{725}{}
\Yearpublication{2011}
\Yearsubmission{2010}
\Month{1}
\Volume{332}
\Issue{1}
\DOI{10.1002/asna.200811027}

\begin{document}

\title{Dynamical quenching with non-local $\alpha$ and downward pumping}
\authorrunning{A. Brandenburg, A. Hubbard, \& P. J. K\"apyl\"a}
\author{
A. Brandenburg\thanks{Corresponding author: brandenb@nordita.org}$^{1,2}$,
A. Hubbard$^{3,1}$,
\& P. J. K\"apyl\"a$^{4,5,1}$}
\institute{
$^1$Nordita, KTH Royal Institute of Technology and Stockholm University,
SE-10691 Stockholm, Sweden\\
$^2$Department of Astronomy, Stockholm University, Roslagstullsbacken 23,
SE-10691 Stockholm, Sweden\\
$^3$Department of Astrophysics, American Museum of Natural History, New York,
NY 10024-5192, USA\\
$^4$Department of Physics, Gustaf H\"allstr\"omin katu 2a
(PO Box 64), FI-00014 University of Helsinki, Finland\\
$^5$
ReSoLVE Centre of Excellence, Department of Information and Computer 
Science, Aalto University, PO Box 15400, FI-00076 Aalto, Finland
}

\date{\today,~ $ $Revision: 1.51 $ $}

\keywords{magnetic fields -- magnetohydrodynamics (MHD)} 

\abstract{%
In light of new results,
the one-dimensional mean-field dynamo model of Brandenburg \& K\"apyl\"a
(2007) with dynamical quenching and a nonlocal
Babcock--Leighton $\alpha$ effect is re-examined for the solar dynamo.
We extend the one-dimensional model to include the effects
of turbulent downward pumping (Kitchatinov \& Olemskoy 2011),
and to combine dynamical quenching with shear.
We use both the conventional dynamical quenching model of Kleeorin \&
Ruzmaikin (1982) and the alternate one of Hubbard \& Brandenburg (2011),
and confirm that with varying levels of non-locality in the $\alpha$ effect,
and possibly shear as well, the saturation field strength
can be independent of the magnetic Reynolds number.
\keywords{MHD -- Turbulence}}

\maketitle

\section{Introduction}

The generation of large-scale magnetic fields is usually explained in terms
of mean-field theory, in which one considers solutions of the averaged
induction equation (Parker 1979; Krause \& R\"adler 1980).
In this theory, the evolution of the mean magnetic field is governed by
turbulent transport coefficients such as the $\alpha$ effect and the
turbulent magnetic diffusivity $\etat$.
As long as the magnetic field is small compared with the equipartition
field strength $\Beq$, and if there is also helicity in the system,
one expects the parameter $\alpha$, which drives dynamo action,
to be of the order of the rms velocity of the turbulence,
and $\etat$ to be of the order of the rms velocity times the mixing
or correlation length of the turbulence (Moffatt 1978).
If this is indeed true, the relevant time scale of the problem should
be the dynamical time scale, rather than the microscopic diffusion time
which would be longer than the dynamical one by a factor that is equal
to the magnetic Reynolds number $\Rm$, which in turn
is very large in many systems of
astrophysical relevance; $10^6$ to $10^9$ in the Solar convection zones.

Early attempts to determine $\alpha$ and $\etat$ from simulations
have suggested that this may not be so simple, 
and that the saturated field strength might decrease rapidly with
increasing $\Rm$ in a phenomenon called ``catastrophic quenching''
(Cattaneo \& Vainshtein 1991; Cattaneo \& Hughes 1996).
The reason for this is that magnetic helicity,
which measures the twist of magnetic flux bundles,
obeys a conservation equation (Gruzinov \& Diamond 1994).
So, as the physics of the $\alpha$ effect describes the twisting of the large-scale
magnetic field by helical fluid motions, it is constrained by
the magnetic helicity equation
in a fashion which resists further twisting of the field.
This is possible because magnetic helicity is signed:
by producing magnetic helicity at small scale of opposite
sign, magnetic helicity can remain constant even while dynamo action
creates helical large scale fields.

In the mean-field formalism, this can be described by an $\alpha$ effect
that depends 
not only on background fluid motions, but
also on the helicity of the small-scale magnetic field.
This provides an extra evolution equation for the magnetic $\alpha$ effect
which describes the production of magnetic helicity locally where strong mean
field twisting occurs.
This approach goes back to the early work of Kleeorin \& Ruzmaikin (1982), and is
now usually referred to as the dynamical quenching formalism.
This formalism has been found to describe many properties of direct
numerical simulations of turbulent dynamos (Field \& Blackman 2002;
Blackman \& Brandenburg 2002; Subramanian 2002).

This formalism is quite different from ``algebraic'' $\alpha$
quenching that is often invoked to describe saturation of the magnetic field
by reducing $\alpha$ locally, depending on the amplitude of the mean field
at that position.
The dynamical quenching formalism can even produce an $\alpha$ effect
where there was none to begin with, for example in the turbulent decay
of a helical large-scale magnetic field (Yousef et al.\ 2003;
Kemel et al.\ 2011; Blackman \& Subramanian 2013).
This can also occur when a mean magnetic field is produced
by the shear--current effect (Rogachevskii \& Kleeorin 2003, 2004).
While the shear--current effect is quite different from the $\alpha$
effect of dynamo theory, 
it produces a helical mean field, and therefore must be accompanied by the 
generation of small-scale
magnetic helicity so that no net magnetic helicity is produced.
This has been demonstrated within the dynamical quenching formalism
(Brandenburg \& Subramanian 2005), where a magnetic $\alpha$ effect
was produced, even though there is no kinetic $\alpha$ effect.
Further examples are the so-called interface dynamos (Parker 1993),
where shear
operates at the bottom of the solar convection zone,
and the kinetic $\alpha$ effect operates at its top.
Again, a magnetic $\alpha$ is produced at locations where strong twisting
of the mean field occurs, regardless of the location of the kinetic
$\alpha$ effect, as was demonstrated by simulations in spherical
geometry (Chatterjee et al.\ 2010).

This situation is similar to models with a Babcock--Leighton $\alpha$
effect which acts at the surface based on magnetic fields at the bottom
of the convection zone (e.g.\ Charbonneau 2010).
This effect is therefore highly non-local.
Dynamical quenching in such a model was considered by
Brandenburg \& K\"apyl\"a (2007; hereafter BK07).
There, dynamical quenching was found to lead to catastrophic
quenching, i.e., the saturation field strength was found to decrease
like $\Rm^{-1}$.
Subsequent work of Kitchatinov \& Olemskoy (2011, 2012) has now shown that,
using a more realistic model of the solar dynamo, catastrophic quenching
may be alleviated in the presence of strong downward pumping.
An alternate new line of research has shown that the ``standard'' set of
dynamical quenching equations can fail in the presence of shear
(Hubbard \& Brandenburg 2011).
The purpose of the present paper is to examine both recent results
in the context of the idealized model of BK07, determining both whether
the results of KO11 depend on a more complicated geometry and
how more recent formulations of dynamical alpha quenching behave
in the presence of non-local phenomena.

\section{Dynamical $\alpha$ quenching and non-locality}

In mean field theory we decompose the fields into mean (overbar) and fluctuating
(lower case) quantities, so for example the magnetic field can be written as
\EQ
\BB = \meanBB + \bb,
\EN
where $\overline{\bb}=0$.  Defining the mean turbulent electromotive
force
\EQ
\meanEMF \equiv \overline{\uu \times \bb},
\EN
we can write the mean field induction equation
(Parker 1979; Krause \& R\"adler 1980)
\EQ
{\partial\meanBB\over\partial t}
=\nab\times\left(\meanUU\times\meanBB+\meanEMF-\eta\mu_0\meanJJ\right).
\label{DynEqn}
\EN
However, if there is helicity in the system, there is
also the occurrence of a magnetic $\alpha$ effect,
$\alpha_{\rm M}$, which characterizes the production of internal
twist in the system and is governed by
\EQ
{\partial\alpha_{\rm M}\over\partial t}+\nab\cdot\meanFF
=-2\eta_{\rm t}k_{\rm f}^2\left({\meanEMF\cdot\meanBB\over B_{\rm eq}^2}
+{\alpha_{\rm M}\over R_{\rm m}}\right),
\label{QuenEqn}
\EN
as described in (Brandenburg \& Subramanian 2005).
$\meanFF$ is the mean flux of small scale magnetic helicity.

In this paper, $\meanEMF$ is assumed to have contributions from the kinetic
and magnetic $\alpha$ effects, $\alphaK$ and $\alphaM$, respectively,
the turbulent magnetic diffusivity $\etat$,
and the turbulent pumping or $\gamma$ effect, i.e., we write
\EQ
\meanEMF=\hat\alphaK\circ\meanBB+\alphaM\meanBB+\ggamma\times\meanBB-\etat\meanJJ.
\EN
We are studying the effect of a non-local BabcockÐ Leighton type $\alpha$,
which generates an $\meanEMF$ that is restricted to the surface layers of the Sun, but depends 
only on the mean magnetic field deep within the convective zone.
Therefore, we treat the kinetic $\alpha$ effect as nonlocal integral kernel,
$\hat\alphaK$, and
\EQ
\hat\alphaK\circ\meanBB=\int_{z_1}^{z_2}\hat\alphaK(z,z')\meanBB(z',t)\,\dd z'
\EN
denotes a convolution, which is here restricted to be only in $z$.

For simplicity, we use a Cartesian domain, with the $xy$ plane
corresponding to surfaces of constant radius in the Sun, and $z$
corresponding to the radial direction.
We have restricted ourselves to $xy$ averages in the
Cartesian domain, so $\meanBB=\meanBB(z,t)$ depends only on $z$ and $t$,
and we assume that the turbulent pumping parameter $\ggamma=\gamma\zzz$
only acts in the vertical direction.
Note that the magnetic helicity equation is unaffected by the
$\gamma$ effect -- just like the large-scale velocity term,
$\meanUU\times\meanBB$, 
it does not directly affect the evolution of $\alphaM$.

\begin{figure}[t!]\begin{center}
\includegraphics[width=\columnwidth]{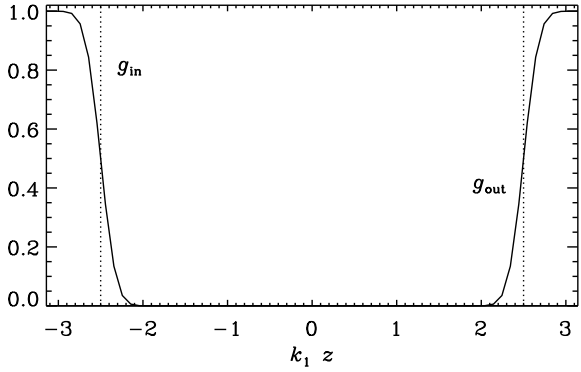}
\end{center}\caption[]{
Contributions $g_{\rm in}(z)$ and $g_{\rm out}(z)$ entering the
nonlocal $\alpha$ effect defined in \Eq{alpkernel}.
The dotted lines indicate the positions of $z_1$ and $z_2$.
}\label{pprofs}\end{figure}

The possibility of nonlocal $\alpha$ and $\etat$ effects has been inferred
also from simulations of magneto-rotational turbulence in accretion discs
(Brandenburg \& Sokoloff 2002) and for turbulence (Brandenburg et al.\ 2008).
In principle, $\etat$ and $\gamma$ should of course also be nonlocal,
but this will here be neglected.
Following BK07, we restrict ourselves to a simple expression of the form
\EQ
\hat\alphaK(z,z')=\alpha_0\, g_{\rm out}(z)\, g_{\rm in}(z'),
\label{alpkernel}
\EN
where $\alpha_0$ is a coefficient, to be specified below, and
\EQ
g_{\rm out}(z)=\half\left[1+\erf\left({z-z_2\over d}\right)\right],
\EN
\EQ
g_{\rm in}(z')=\half\left[1-\erf\left({z'-z_1\over d}\right)\right]
\EN
are simple profile functions representing the peak of the source function
near $z=z_2$ with a sensitivity for fields located near $z=z_1$.
For the following we choose $-z_1=z_2=2.5/k_1$ and $d=0.05/k_1$ in the domain
$-\pi<k_1 z<\pi$; see \Fig{pprofs}.
Here, $k_1$ is the smallest wavenumber in the computational domain
and is used as our inverse unit length.

In most of the published literature on dynamical quenching,
the magnetic helicity flux has no contribution from a term
$\meanEMF\times\meanAA$, which enters with opposite signs in
the evolution equations for the magnetic helicity flux from
small-scale and large-scale fields.
However, recent work (Hubbard \& Brandenburg 2011,2012) now reveals
that this is not permissible, and including it tends to
alleviate catastrophic quenching.
Nonetheless, for comparison with the published literature,
we solve \Eqs{DynEqn}{QuenEqn} first for the case $\meanFF=0$.
We use an implicit scheme for $\alpha_{\rm M}$, as described in BK07.
We have considered a model using linear shear of the form
$\meanUU(z)=(0,S(z) x,0)$.
In this paper, $\gamma$ and $\etat$ are assumed constant.
The strength of shear, $\alpha$, and $\gamma$ effects is quantified by the
non-dimensional numbers
\EQ
C_S={S\over\eta_{\rm t}k_1^2},\quad
C_\alpha={\alpha_0\over\eta_{\rm t}k_1},\quad
C_\gamma={\gamma\over\eta_{\rm t}k_1}.
\label{Calp_etc}
\EN
In the following we use $C_S=100$, $C_\alpha=0.1$, and $k_{\rm f}/k_1=5$,
while $C_\gamma$ will be varied.
In many cases an explicit treatment of the $\alphaM$ equation
suffices (e.g.\ Blackman \& Brandenburg 2002), but in the present case
an explicit solution algorithm was found to be unstable; see BK07 for details.

\begin{figure}[t!]\begin{center}
\includegraphics[width=\columnwidth]{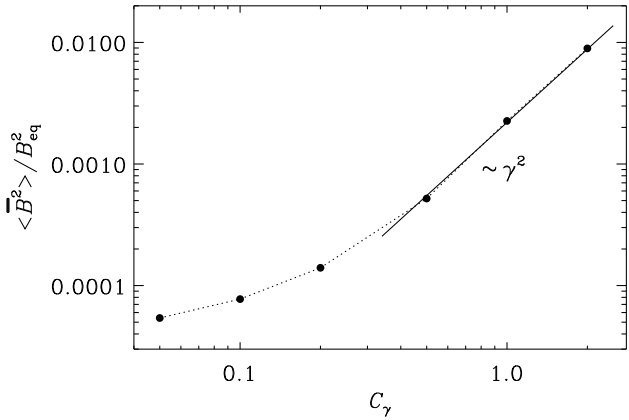}
\end{center}\caption[]{ 
Dependence of the saturation field strength on $C_\gamma$
for $\Rm=10^4$.
}\label{pgam_dep}\end{figure}

\subsection{Homogeneous shear}

\begin{figure}[t!]\begin{center}
\includegraphics[width=\columnwidth]{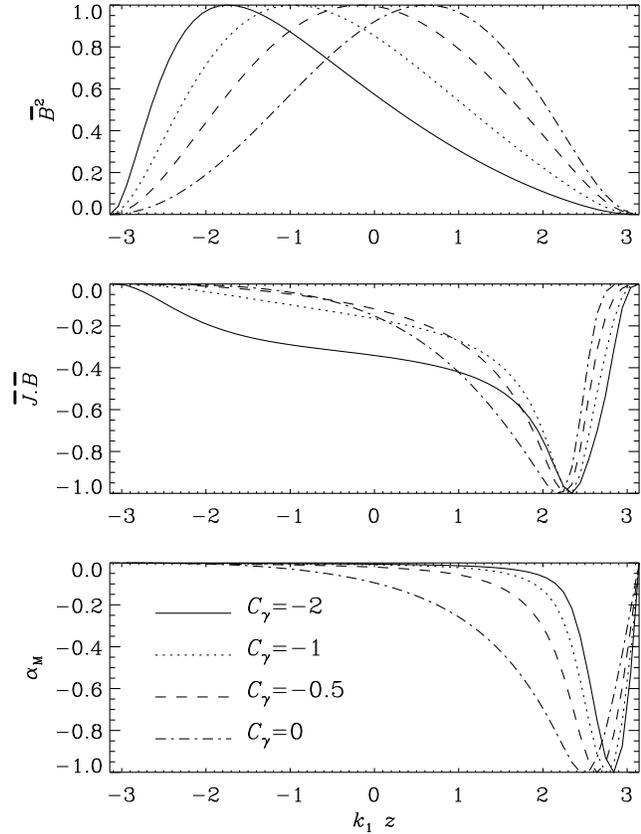}
\end{center}\caption[]{ 
Vertical dependence of $\meanBB^2$, $\meanJJ\cdot\meanBB$, and $\alphaM$ for a model
with homogeneous shear,
all normalized by their local extrema, for different values of $C_\gamma$.
}\label{pbb_gam}\end{figure}

\begin{figure}[t!]\begin{center}
\includegraphics[width=\columnwidth]{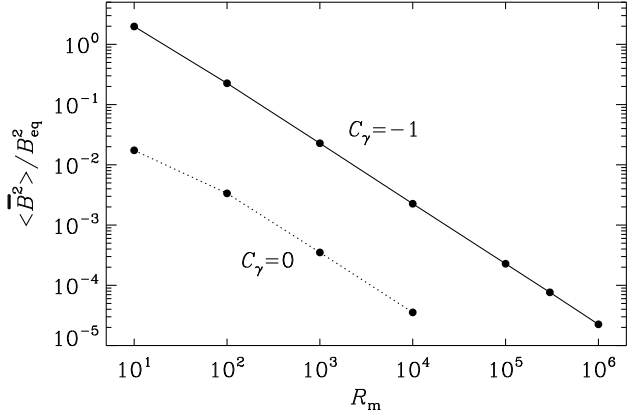}
\end{center}\caption[]{ 
Dependence of the saturation field strength on $\Rm$
for models with homogeneous shear, with and without downward pumping or $\gamma$ effect.
}\label{pRm_dep_wgam}\end{figure}

While the Sun has a strong shear layer at the base of the convective zone,
we first consider the case of homogeneous shear 
($C_S$ not depending on $z$).
In this case, it turns out that the inclusion of downward pumping
in the model of BK07 makes the dynamo stronger, as can
be seen in \Fig{pgam_dep}, where we show that $\meanBB^2\propto\gamma^2$.
The reason for this is that most of the field is generated in the middle of the
domain, while most of the quenching via $\alphaM$ occurs near
the top of the layer around $z=z_2$; see \Fig{pbb_gam}.
Nevertheless, this model still experiences catastrophic
quenching; see \Fig{pRm_dep_wgam}.
These results are quite similar to those obtained for local $\alpha$
profiles (Brandenburg \& Subramanian 2005).
We note that for $R_{\rm m}>10^4$ it is important to perform the
calculations using double precision arithmetics.

A somewhat surprising property of the present solutions is the fact
that $\meanJJ\cdot\meanBB$ is still negative everywhere; see the middle panel
of \Fig{pbb_gam}.
This is mainly a consequence of the nonlocal $\alpha$ effect;
for a local $\alpha$ effect, and certainly in the absence of shear,
$\meanJJ\cdot\meanBB$ would always be positive for positive $\alpha$.
Nevertheless, $\alphaM$ is negative everywhere, 
the opposite sign as the kinetic $\alpha$ effect, so there is no
possibility of having anti-quenching anywhere in the domain.

\begin{figure}[t!]\begin{center}
\includegraphics[width=\columnwidth]{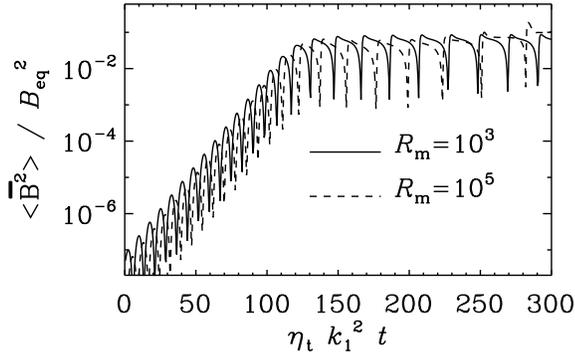}
\end{center}\caption[]{ 
Saturation behavior of a dynamo with a shear profile that is
identical to that of $g_{\rm in}$, which is localized
only to the lower layer.
}\label{pcomp_shear}\end{figure}

\subsection{Shear layer}

Next, inspired by the solar tachocline, we consider a model where
shear is confined to a narrow layer at the bottom of the domain.
In that case we replace $S$ by $S(z)=S_0 g_{\rm in}$, i.e., the shear
layer coincides with the profile with which the $\alpha$ kernel operates
on the magnetic field.
It turns out that in that case the magnetic field becomes oscillatory.
This means that $\alphaM$ can now change sign and thereby offset
the quenching such that the saturation level becomes independent of
the magnetic field strength.
This is shown in \Fig{pcomp_shear}, where we plot $\bra{\meanBB^2}$
versus time for two values of $\Rm$.
Both the linear growth rate and the initial saturation field strength
are now found to be independent of $\Rm$.
(For smaller values of $\Rm$ the linear growth rate would become
progressively smaller, because the effective dynamo number would decrease.)
For $\Rm=10^3$ the model saturates at a fixed level for all times, but
in models with larger values of $\Rm$ ($10^4$ or $10^5$) the field achieves
only temporary saturation before it grows beyond any limit.
This is in agreement with earlier models of Kitchatinov \& Olemskoy (2011)
and other dynamical quenching models, for example in models of
Brandenburg \& Subramanian  (2005), in which an $\alphaM$ is driven by a
magnetic helicity flux of Vishniac--Cho type (Vishniac \& Cho 2001).

\section{Alternate quenching}

\begin{figure}[t!]\begin{center}
\includegraphics[width=\columnwidth]{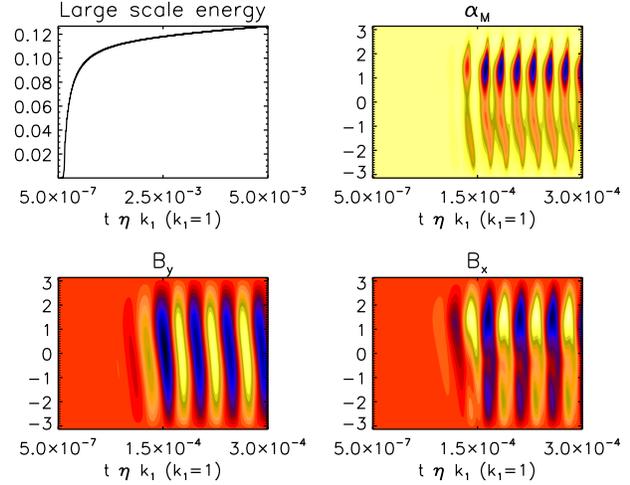}
\end{center}\caption[]{ 
Dynamo behavior at early times with homogeneous shear
using the alternate quenching formulation.
$C_S=-100$, $C_{\alpha}=-0.1$, $C_{\gamma}=-1$, $\Rm=10^5$.
Note the difference in the scale of the $x$-axes, the dynamo waves
are too numerous to plot fully. Further, the energy in the top left panel
is a running mean.
}\label{long1}\end{figure}

Quite different results are obtained when a different formulation
of dynamical $\alpha$ quenching is used.
Using the results in Hubbard \& Brandenburg (2011,2012), we can replace
\Eq{QuenEqn} with
\begin{eqnarray}
&&\partial \overline{h}/\partial t=-2 \eta \left(\meanJJ \cdot \meanBB
       +\alpha_M \Beq^2/\etat\right) \\
&& \alpha_M=\etat \kf^2 \left(\overline{h}-\meanAA \cdot \meanBB\right)/\Beq^2.
\end{eqnarray}
This formulation gives better results in the geometries studied in
Hubbard \& Brandenburg (2012),
namely shearing-periodic with a homogeneous $\alpha$ effect.
This approach has now also been applied to solar-like models in
spherical geometry (Pipin et al.\ 2013).

In this case, it is important to consider how an oscillatory dynamo functions.
First the $x$-component of the field at the bottom, $B_x^{\rm bot}$,
is sheared into a $y$-directed field $B_y^{\rm bot}$
with $\sgn B_y^{\rm bot}=\sgn S B_x^{\rm bot}$.
In the assumed high-shear regime ($|C_S| \gg C_\alpha, C_\gamma$),
this $y$ directed toroidal field dominates the energetics.
From this toroidal field the non-local $\alpha$-effect generates an $x$-directed
poloidal field $B_x^{\rm top}$ at the top, with
$\sgn B_x^{\rm top}=\sgn(-\alpha' S B_x^{\rm bot})$,
where the prime denotes a $z$ derivative due to taking the curl in \Eq{DynEqn}.
If $\sgn B_x^{\rm top} \neq \sgn B_x^{\rm bot}$, then when the field is
transported downwards (via pumping through the gamma effect, or diffusion),
it will counter the
original field, resulting in dynamo waves, as seen in \Fig{long1}.
If the signs are the same, there is only amplification, with no
back-reaction mechanism available in the formalism we consider,
although algebraic quenching or similar must eventually play a significant
role.
In this section we therefore only present results for the oscillatory case.

\begin{figure}[t!]\begin{center}
\includegraphics[width=\columnwidth]{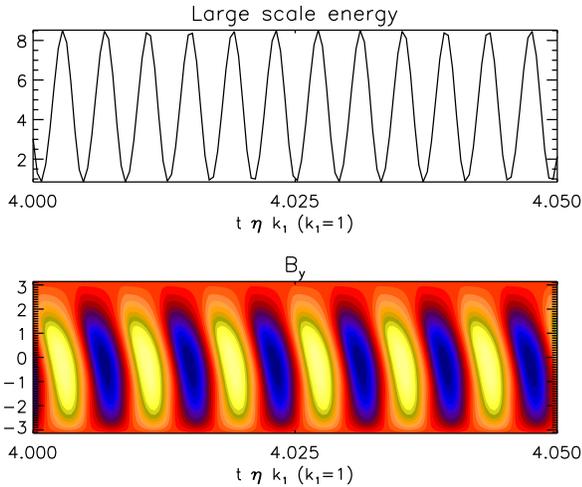}
\end{center}\caption[]{ 
Dynamo behavior at intermediate times with homogeneous shear for
a dynamo with
$C_S=-100$, $C_{\alpha}=-0.1$, $C_\gamma=-1$, $\Rm=500$.
The alternate quenching formulation was used.
Here the energy is not a running mean.
}\label{Energy}\end{figure}

This nonlocal $\alpha\Omega$ dynamo does not have the same behavior as
a uniform one.
Perhaps most significantly, the energy in the magnetic fields
shows strong fluctuations as the magnetic field oscillates.
Accordingly, for most figures in this
section, our energies are running means.
The exception is \Fig{Energy}, where we show a time-magnification of the
dynamo, and strong time variation is visible.

\begin{figure}[t!]\begin{center}
\includegraphics[width=\columnwidth]{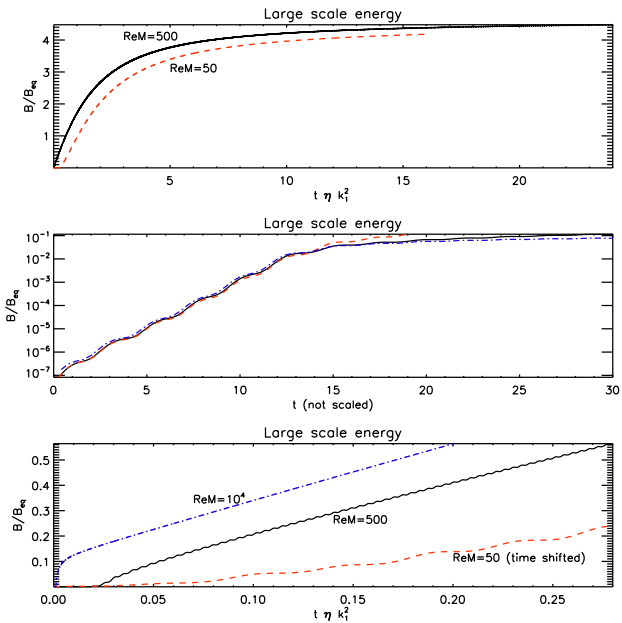}
\end{center}\caption[]{ 
Alternate quenching dynamo behaviors for three values of $\Rm$,
other parameters $S=-100$, $\alpha=-0.1$, $\gamma=-1$.
Black/solid: $\Rm=500$; red/dashed: $\Rm=50$;
blue/dash-dotted: $\Rm=10^4$ (not included in top panel).
}\label{ptriple}\end{figure}

In \Fig{ptriple} we examine the dependency of the system on $\Rm$.
In the top panel, we see that the time dependence is similar
for $\Rm=50,\ 500$, although it appears that $\Rm=50$ is
not yet in the asymptotically high $\Rm$ regime.
Note that the time axis is scaled to the resistive time.
In the middle panel, we show the early (kinematic) behavior,
which is identical for all three values of $\Rm=50,\ 500,\ 10^4$,
with a non-scaled time axis.
In the bottom panel, we see the early non-linear evolution,
where the results for $\Rm=500,\ 10^4$ are identical, but again it
appears that $\Rm=50$ is too low for fully asymptotic behavior.  Note
that the time of entrance into the linear growth regime is different
for the three cases because of the scaling of $t$.
Intriguingly, the slow-saturation phase shown is similar to that predicted
for closed helical systems.  We see no evidence for a declining final field
strength with $\Rm$.

\begin{figure}[t!]\begin{center}
\includegraphics[width=\columnwidth]{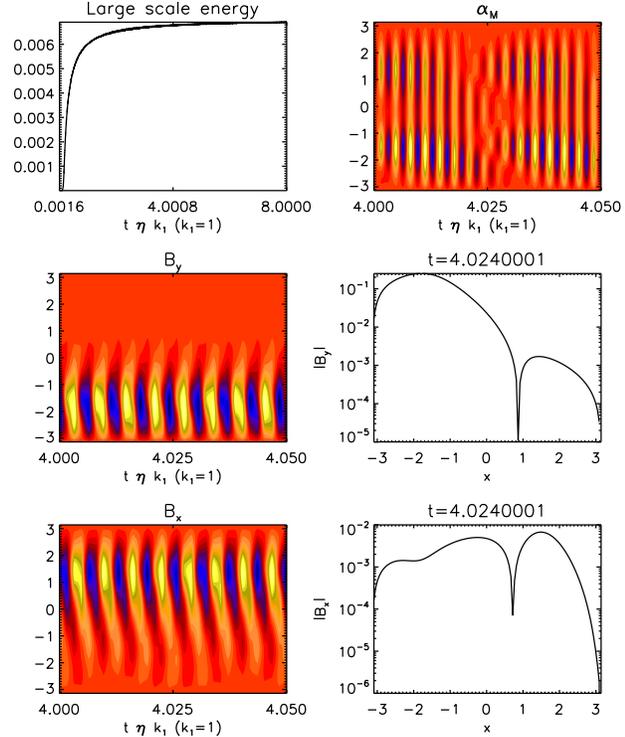}
\end{center}\caption[]{ 
Alternate quenching dynamo behavior at intermediate times with
inhomogeneous shear.
$C_S=-100$, $C_{\alpha}=-0.2$, $C_\gamma=-1$, $\Rm=500$.
Notably, here the shear is not homogeneous, instead it has the
same spatial variation as $g_{\rm in}$.  The dynamo control
parameters had to be changed ($C_{\alpha}$ from $-0.1$ to $-0.2$) to
allow dynamo growth.  The dips in the temporal log-linear cuts
are sign-changes.
}\label{nlS}\end{figure}

As a final comment in this section, in \Fig{nlS} we include a run with
not merely a non-local $\alpha$ effect, but also
non-local shear, to model a Babcock-Leighton $\alpha$ effect at the surface,
and the strong shear localized in the solar tachocline.
The dynamo is overall similar to the case with homogeneous shear
but, predictably, far less strongly
excited, necessitating an increased magnitude of $C_\alpha=-0.2$

\section{Conclusions}

The present work has demonstrated that, within the framework of the
dynamical quenching model, a nonlocal $\alpha$ effect of Babcock-Leighton
type combined with downward pumping can alleviate catastrophic quenching
only when the shear layer is separated from the layer where the
Babcock-Leighton $\alpha$ acts.
Downward pumping can lead to a strong enhancement of the dynamo
even in models where shear is uniform.
While this can compensate for some of the field reduction suffered
from large values of $\Rm$, it nevertheless does not change the $\Rm^{-1}$ scaling.

The model of Kitchatinov \& Olemskoy (2011) contains an important
feature that is found here only in time-dependent cases, namely the
sign reversal of $\alphaM$ in some
places, which leads to catastrophic anti-quenching (or amplification).
Similar sign reversals of the
local value of $\alphaM$ have been seen in some earlier models with a
local $\alpha$ effect (Guerrero et al.\ 2010; Chatterjee et al.\ 2011),
but it needs to be seen whether this behavior is physically realistic and still
compatible with the original equations.

Further,
it is clearly only a simplification to neglect the flux term in \Eq{QuenEqn}.
Even though the domain may be closed, we must always expect there to be
internal magnetic helicity fluxes resulting from the inhomogeneity of
the model.
Magnetic helicity fluxes between local extrema in the small-scale
magnetic helicity density and across the equator have been detected in
direct numerical simulations (Mitra et al.\ 2010;
Hubbard \& Brandenburg 2010; Del Sordo et al.\ 2013).
Such fluxes might well be sufficient for alleviating catastrophic quenching
without the need for invoking the non-locality of $\alpha$.

On the other hand, an improved integration of shear with dynamical quenching
can avoid catastrophic $\alpha$-quenching, but only functions and
generates an oscillatory dynamo when the signs of the $\alpha$
effect and the shear are the same.  When the signs are different, dynamical
quenching predicts no feedback, so the field grows without bound (although
some form of geometric $\alpha$ quenching must eventually control the system).

\acknowledgements
We thank Kandaswamy Subramanian for comments and encouragement.
This work was supported in part by the Swedish Research Council,
grants 621-2007-4064 and 2012-5797 (AB), the European Research Council
under the AstroDyn Research Project 227952 (AH), and the
Academy of Finland grants No.\ 136189, 140970, 272786 (PJK).
AH received additional support from the Alexander von Humboldt Foundation
and NASA OSS grant NNX14AJ56G.


\end{document}